\documentstyle[12pt,epsfig]{article}
\begin{document}
\noindent
\begin{center}
{\Large {\bf  On Non-triviality of Units Transformations}}\\
\vspace{2cm}
 ${\bf Yousef~Bisabr}$\footnote{e-mail:~y-bisabr@sru.ac.ir.}\\
\vspace{0.5cm} {\small{Department of Physics, Shahid Rajaee Teacher
Training University,
Lavizan, Tehran 16788, Iran}}\\
\end{center}
\vspace{1cm}
\begin{abstract}
There are some theoretical arguments about possible variations of fundamental constants with cosmic time. We review the fact that all conversion factors depend on these quantities and consider how their variations may affect transformations of units. We deal with the relevance of the issue to the cosmological constant problem.

\end{abstract}

\section{Introduction}
The main objective of physics is to discover the laws by which Nature works through experimental observations and quantitative measurements. Thus measurement is an integral part of physical theories and without ability to measure, it would be impossible for physicists to construct theories.
In a measurement process, numerical values of physical quantities are compared with some reference values which are used as units, e.g., $1$ meter or $2$ seconds. This requires us to define appropriate units systems. There are different units systems which are usually used in daily life and scientific works. In a particular measurement, some of these units systems are more suitable than the others. For instance, we would like to use Angstrom to measure the radius of electron orbit in an atomic system rather than light-year while the latter is more suitable to measure the distance of the sun to the center of the Milky Way.\\  Due to certain purposes, we usually need to give the result of a measurement in different units. In these cases, we use a unit transformation which is a conversion between different units for the same quantity. This is done by using appropriate multiplicative conversion factors. Conversion factors are generally taken to be constant in space and time. After all, inch should be converted to millimeter regardless of where and when it is done. Such unit transformations are trivial in the sense that they are devoid of any dynamical meaning and different units are actually constant multiple of each other.\\
Here we intend to pose a debate on this triviality. We consider the origins of such a debate and argue that they
are related to the fact that definition of a system of units is very tied-up to the concept of fundamental physical constants. On the other hand, there are theoretical arguments in favour of variations of these 'constants' \cite{co}. The consequence of such variations is non-constancy of conversion factors and then non-triviality of a unit transformation. As an illustration of possible impacts of such a non-triviality on physical theories, we shall deal with the cosmological constant problem \cite{bound} \cite{ccc}.
~~~~~~~~~~~~~~~~~~~~~~~~~~~~~~~~~~~~~~~~~~~~~~~~~~~~~~~~~~~~~~~~~~~~~~~~~~~~~~~~~~~~~~~~~~~~~~~~~~~~~~~~~~~~~~~~~~~~~~~~~~~~~~~~

\section{Units systems and fundamental constants}
The first step towards describing natural phenomena is to introduce various entities such as fields and constants on which those phenomena may be conceived to depend. The next step is to discover relations among these entities and to construct a mathematical model to explain the phenomena and make future predictions. The last step is to determine the degree to which those predictions are accurate representations of the real world.  For instance, electromagnetic phenomena are described by Maxwell's theory which introduces electric and magnetic fields. The theory contains equations that explain relationships between electricity and magnetism and correctly predict that these fields propagate as waves with
a constant velocity.\\
Measurement of physical quantities in different units is the major part of the process of model building. Evidently, the use of a particular unit of measure is a matter of convenience and depends on the intended purposes. One can change the unit of a dimensional quantity by applying a convenient conversion
 factor. The conversion factor is a ratio which converts the unit of that quantity into another one without changing its dimension.\\
 It is possible to divide all units systems into two  groups: In the first group (\textbf{GI}), there are all units which are used in daily life, commerce and also engineering. It consists of MKS, CGS or the Imperial  units.
The second group (\textbf{GII}), contains all units that are used in physical theories. They are defined in terms of the so-called fundamental  constants (FCs). Fundamental theories of physics introduce universal quantities which are usually considered to be constant in space and time. They have essential roles in theories and experiments but they can not be calculated even though it is possible to determine their numerical values via accurate measurements.
Some examples of these FCs are $c$ (velocity of light) and $e$ (electric charge) in electromagnetic theory; $h$ (Planck constant), $m_e$ and $m_p$ (the rest masses of electron and proton, respectively) in quantum theories and $G$ (gravitational constant) in gravitational theories. Some of the scientific units and the corresponding conversion factors are indicated in the tables $1$\footnote{The Hartree units is a system of natural units which is useful in measurements of atomic physics. It is named after the English mathematician and physicist Douglas R. Hartree (1897-1958).} and $2$.

\begin{table}[h]
\caption{Some popular scientific units for length, time and mass. }
\centering
\begin{tabular}{|c| c| c| c| }
\hline\hline

quantity &Hartree   & quantum  & Planck \\
$$ & atomic(a) units &  (q) units &(p) units\\

\hline
length(L) & $L_{a}=\frac{\hbar^2 }{m_e e^2}$ &
$L_{q}=\frac{\hbar}{m_p c}$ & $L_p=\sqrt{\frac{\hbar G}{c^3}}$ \\

 \hline

time(T) & $T_{a}=\frac{\hbar^3 }{m_e e^4}$ &
$T_{q}=\frac{\hbar}{m_p c^2}$ & $T_p=\sqrt{\frac{\hbar G}{c^5}}$\\
\hline

mass(M) &  $M_{a}=m_e$ &
$M_{q}=m_p$ &
$M_p=\sqrt{\frac{\hbar c}{G}}$ \\
\hline
\end{tabular}
\label{1}
\end{table}

\begin{table}[h]
\caption{Some conversions of the table $1$. }
\centering
\begin{tabular}{|c|c|}
\hline\hline
$ a \rightarrow q$ &
$q\rightarrow p$   \\
\hline
$\frac{e^2 m_e}{\hbar c m_p}$&
$\frac{m_p\sqrt{G}}{\sqrt{\hbar c}}$\\
 \hline
$\frac{e^4 m_e }{\hbar^2 c^2 m_p} $&
 $\frac{m_p\sqrt{G}}{\sqrt{\hbar c}}$ \\
 \hline
$\frac{m_p }{ m_e }$ &
$\frac{\sqrt{\hbar c}}{m_p\sqrt{G}}$ \\
\hline
\end{tabular}
\label{2}
\end{table}
We define a unit transformation to be trivial when the corresponding conversion factor is independent of space and time.
Such a triviality requires an implicit assumption concerning constancy of FCs. Although there has been no confirmed observational evidence yet against this assumption, we will argue in the next section that there are compelling theoretical reasons for expecting variations in a cosmological time scale\footnote{Although FCs may generally vary with both space and time, we consider such variations
with cosmic time.}.

~~~~~~~~~~~~~~~~~~~~~~~~~~~~~~~~~~~~~~~~~~~~~~~~~~~~~~~~~~~~~~~~~~~~~~~~~~~~~~~~~~~~~~~~~~~~~~~~~~~~~~~~~~~~~~~~~~~~~~~~~~~~~~~~
\section{Are fundamental constants really constant?}
Dirac was one of the first well-known physicists who brought up the subject of possible variations of FCs. He combined them with some cosmological parameters to obtain some dimensionless large numbers with approximately the same orders of magnitudes \cite{dir}. He then argued that the near equality of these quantities represents something more than an accidental coincidence. For instance, combination of fundamental constants in quantum physics with  $G$ gives
\begin{equation}
N_1\equiv \frac{\hbar c}{G m_p^2}\sim 10^{39}
\end{equation}
Moreover, we may obtain numbers with the same orders by inclusion of the Hubble constant $H_0$ and the matter density of the observable Universe $\rho_0$
\begin{equation}
N_2\equiv \frac{m_p c^2}{\hbar H_0}\sim 10^{39}
\end{equation}
\begin{equation}
N_3\equiv \frac{\rho_0 c^3}{m_p H_0^3}\sim 10^{79}
\end{equation}
Note that $H_0$ is the present value of the expansion rate $H$ and the latter evolves with cosmic time. Dirac argued that the relation $N_1\simeq N_2\simeq\sqrt{N_3}$ is not just a coincidence taking place at the present stage of the Universe but instead it should be held true during evolution of the Universe. This implies that $N_2$ and $N_3$ evolve with the expansion and at least one of the constants used in $N_1$ is not a genuine constant and changes with time. In his original model, Dirac proposed that $G$ decreases with time as $1/t$. Later, it was shown that this proposal was not consistent with astrophysical and cosmological bounds on variation of $G$ \cite{cont}.
Dirac's arguments are indeed phenomenological analyses that create new insights into investigations of physical constants. His viewpoint subsequently led to the establishment of physical theories in which the FCs (specifically, $G$ or the fine-structure constant) become dynamical fields which vary with space and time and satisfy well-defined field equations \cite{jor}.\\
Motivations for variability of FCs also appear within theories which assume extra dimensions for space. Examples are Kaluza-Klein and string theories which search for unification of gravity with electromagnetism and quantum physics, respectively, in higher dimensional spacetimes. A general feature of these theories is that the traditional constants of nature are genuine constants only in the full higher dimensional theories. In effective four-dimensional versions, obtained via some processes of the so-called dimensional reduction, they vary with space and time. In the four-dimensional world, those 'constants' depend actually on some dynamical scalar fields such as dilatons and moduli whose dynamics are determined by the theory.\\
If FCs are allowed to vary, basic units and then measurement devices such as clocks and rulers will also be spacetime dependent. To see this, let us look at the definitions of the basic units of the MKS system. The meter is defined as the length of the path traveled by light in vacuum during $1/299,792,458$ of a second. The second is defined in terms of the frequency of
the transition between two hyperfine levels of the ground state of the cesium-$133$ atom. The kilogram is defined by taking the fixed numerical value of the Planck constant to be $h=6.62607015\times10^{-34}$ $Kg~ m^2~ s^{-1}$ \cite{kg}.\\
From an observational point of view, there have been many attempts trying to detect possible variations of FCs with space or time. Some of them are laboratory-based experiments and others rely on astrophysical or cosmological observations \cite{uzan}. Recently, there have been observational evidences providing suggestions that the fine structure constant $\alpha\equiv e^2/\hbar c$\footnote{In Gaussian units in which $4\pi \epsilon_0=1$.}, the measure of the strength of the electromagnetic interaction between photons and electrons, and the electron-proton mass ratio $\mu\equiv m_e/m_p$ might have varied slightly over cosmological time \cite{ba}. Webb \emph{et al.} \cite{web} found
$\Delta\alpha/\alpha\equiv\alpha(z)-\alpha(0)/\alpha(0)=(-0.57\pm 0.10)\times 10^{-5}$ between redshift $z$ and the present day ($z=0$) using quasars absorbtion spectral lines. Moreover, Reinhold \emph{et al.} \cite{rei} reported an indication of a variation $\Delta\mu/\mu=(-24.4\pm 5.9)\times 10^{-6}$ using highly accurate laboratory-based measurements. Unfortunately, later reanalysis of those experiments and the systematic errors casted doubts on validity of these results \cite{doubt}.\\
The dependence of metrology on FCs raises a question about how sensible is it to speak of variation of a dimensional quantity? In fact, there is an inherent ambiguity related to this subject. Measuring variations of a dimensional quantity (such as a FC) needs a units system that itself is defined in terms of FCs. Definition of units
systems is so entangled with FCs that it is only meaningful to consider
variations of dimensionless quantities. However, daily life and engineering are full of dimensional
quantities which we need to measure. How meaningful is to speak of measuring masses,
durations of time intervals, velocities and so on? To answer this question one should bear in
mind that variations of FCs are supposed to take place in a time scale comparable to the age
of the Universe. In human time scales, the supposed variations
are so small that they can certainly be neglected and FCs effectively remain constant.\\
In conclusion, all theoretical arguments that consider possibility of variations of FCs also predict
non-triviality of both \textbf{GI} and \textbf{GII}. Even though \textbf{GI} can be roughly regarded as trivial in human time scales, we shall argue in the next section that disregarding the non-triviality of \textbf{GII}
may have serious consequences.
~~~~~~~~~~~~~~~~~~~~~~~~~~~~~~~~~~~~~~~~~~~~~~~~~~~~~~~~~~~~~~~~~~~~~~~~~~~~~~~~~~~~~~~~~~~~~~~~~~~~~~~~~~~~~~~~~~~~~~~~~~
\section{What is the relevance of such a non-triviality?}
An immediate deduction of non-triviality of a units transformation is that such a transformation should be regarded as a dynamical process rather than a naive multiplication by a constant factor. In this case, the conversion relating the two units is no longer a constant and should be taken as a dynamical field satisfying a dynamical field equation in a field theory context. In this section we review the cosmological constant problem ($\Lambda$-problem) and highlight its relevance to this issue.
\subsection{The cosmological constant}
In $1905$, Einstein introduced special theory of relativity in which space and time brought into a unique entity known as spacetime. Ten years later, he introduced general relativity (GR) as a generalization of the special theory. In GR, Einstein provided a refinement of Newton's gravitational theory and interpreted gravity as a geometric property of spacetime. In this theory, the curvature of spacetime in a region is connected to distribution of matter in that region. This connection is characterized by the Einstein field equations\footnote{We use the units system in which $c=1$ and the sign conventions (-,+,+,+). }
\begin{equation}
G_{\mu\nu}=8\pi G T_{\mu\nu}
\label{1}\end{equation}
 The second-rank symmetric tensors $G_{\mu\nu}\equiv R_{\mu\nu}-\frac{1}{2}g_{\mu\nu}R$ and $T_{\mu\nu}$ are the so-called Einstein and energy-momentum tensors, respectively. These equations are $10$ coupled non-linear differential equations relating components of the dynamical tensor field $g_{\mu\nu}$ to physical properties of the matter distribution encoded in $T_{\mu\nu}$. In the case that curvature of spacetime and also velocities of matter particles are sufficiently small (the weak field approximation), these equations are reduced to the Poisson's equation for gravitational potential in Newtonian gravity. Thus the full GR field equations contain all properties of newtonian gravity and also predict relativistic effects such as deviations from the classical precession of Mercury's perihelion experimentally verified in $1916$.\\
Once Einstein had completed his theory, it was natural for him to apply it to the Universe as a whole. He soon realized that
his equations were not consistent with the assumption of a Universe with a static distribution of matter\footnote{Evidences to the contrary viewpoint did not emerge until Hubble's observation was announced in $1929$.}. There was a similar problem in Newtonian gravity. In that case, the Universe
was assumed to be filled with a uniform static
mass density $\rho$ distributed over an infinite Euclidean three-dimensional
space. Each particle in the distribution is subjected to a gravitational potential $\phi\equiv G \int\frac{\rho(r)}{r}dV$ with $dV$ being a volum element.
It is evident that there is no a meaningful definition of $\phi$ at spatial infinity (for a review, see \cite{r} and the references therein.).
In his $1917$ paper entitled 'Cosmological considerations on the general theory of relativity' \cite{e1}, Einstein dealt with the problem and  proposed modifications for Poisson's and his own equations by adding a constant term.
The equations (\ref{1}) took then the form
\begin{equation}
G_{\mu\nu}+\Lambda g_{\mu\nu}=8\pi G T_{\mu\nu}
\label{5}\end{equation}
where $\Lambda$ was known later as the cosmological constant.  Einstein showed that the idea of a static distribution of matter could be preserved by attributing an appropriate value to $\Lambda$ and assuming that the Universe was spatially closed.  After Hubble's observation of expansion of the Universe \cite{hub}, it seemed that the cosmological constant term was unnecessary and thus it was discarded by Einstein.\\
After the development of quantum field theory in the $1940s$, the $\Lambda$-term come back into the game. There was, however, a controversy between theoretical estimations and observational value of $\Lambda$ known as the $\Lambda$-problem.
\subsection{The $\Lambda$-problem}
 Let us consider a Universe devoid of any kind of matter or radiation. In this case, $T_{\mu\nu}=0$ and the equations (\ref{5}) are reduced to $G_{\mu\nu}=-\Lambda g_{\mu\nu}$. Comparing the latter with (\ref{1}) reveals that it is possible to interpret $T_{\mu\nu}^{vac}\equiv -\frac{\Lambda}{8\pi G} g_{\mu\nu}$ as the energy-momentum tensor of vacuum. Following this line, one can take $T_{\mu\nu}^{vac}$ as a perfect fluid with energy density and pressure defined as $\rho^{vac}=-p^{vac}=\frac{\Lambda}{8\pi G}$. Note that such a perfect fluid has a negative pressure. Thus the cosmological constant is equivalent to a repulsive force that acts  opposed to the force of gravity. It was this peculiar behavior of $\Lambda$ that allowed Einstein to counterbalance the gravitational force and to obtain a static Universe.\\
How does one attribute energy density to vacuum?  Here is where one should distinguish between classical and quantum descriptions of vacuum. As a simple example, we consider a particle subjected to a one-dimensional potential of a simple harmonic oscillator $V(x)=\frac{1}{2}\omega x^2$. Classically,
the vacuum for such a system is the state in which the particle is at rest at the
minimum of the potential at $x = 0$. Evidently, this is a state with zero energy. Quantum mechanically, the vacuum is the state with the lowest possible energy. In this case, the uncertainty principle forbids us to associate definite values both to
position and momentum of the particle. One finds that the minimum (ground state) energy of the particle is not zero and is given by $E_0=\frac{1}{2}\hbar\omega$ \cite{qm}.\\
This situation for a particle can be extended to quantum field theory \cite{fe}. A classical field is an entity which is defined through all points of space and time. In momentum space, therefore, a free quantum field can be thought of as a collection of infinite quantum harmonic oscillators oscillating with different energies. The vacuum energy of the field is the sum over all ground state energies of the oscillators which is given by $E=\sum_{i}\frac{1}{2}\hbar\omega_i$ or for energy density
\begin{equation}
\rho^{vac}_{th}=\frac{E}{V}=\frac{1}{V}\sum_{i}\frac{1}{2}\hbar \omega_i
\label{6}\end{equation}
 The result of the sum over infinite number of oscillators is clearly infinite. This is known as 'ultraviolet divergence' since it arises due to contributions of oscillators with very high energies. To give a meaningful expression to the vacuum energy density, we usually do not allow oscillators with arbitrarily large energies to contribute to the sum. The sum is cut off at a particular high energy (or frequency) up to which the field theory under consideration is supposed to be valid\footnote{That energy scale is taken to be the Planck energy below which the predictions of the Standard Model of particle physics, quantum field theory and general relativity are consistent. Beyond that energy, quantum effects of gravity are expected to dominate and thus the above calculation needs to be modified.}. One can do the sum in (\ref{6}) by putting the system in a three dimensional box and letting its volume go to infinity. Applying the method of periodic boundary condition, leads to \cite{bound} \cite{ca}
\begin{equation}
\rho^{vac}_{th}\propto \int_0^{\omega_{max}}\omega^3 d\omega\propto \omega^4_{max}
\label{7}\end{equation}
By taking the Planck energy $E_p$ as the cut off, we expect $\rho^{vac}_{th}\sim (10^{18}Gev)^4\sim 10^{110}erg/cm^3$.\\
There is also an observational value for $\Lambda$ which is provided by measurement of luminosities of sample galaxies in terms of distance or cosmological redshfit. These observations imply $\rho^{vac}_{obs}\sim (10^{-12}Gev)^4\sim 10^{-10}erg/cm^3$. The $\Lambda$-problem is the sharp contradiction between theory and observation which is an incredible $120$ orders of magnitude. This huge discrepancy implies existence of  deep inconsistencies in basic assumptions leading to this result.
\subsection{A reduction mechanism}
There have been many attempts trying to resolve the $\Lambda$-problem \cite{w}. Most of them are based on the belief that $\Lambda$ may not have such an extremely small value suggested by observations
at all the time. In fact, there should exist a dynamical mechanism working during evolution of the
universe which has led to reduction of the large vacuum energy density to a small value. In this viewpoint, one accepts that
the theoretical estimations and the observational value of $\Lambda$ (which are denoted in the following by $\Lambda_{th}$ and $\Lambda_{obs}$) belong to two different epochs of evolution of the Universe. There are at least two reasons why $\Lambda_{th}$ should be related to the early times. The first reason concerns with the fact that all theoretical predictions of $\Lambda$ are based on fluctuations of quantum fields which are important only at early times where energy of the fields are so high that classical descriptions are no longer valid. Furthermore, standard model of particle physics implies that the universe has undergone a series of phase transitions at early epoch of its
evolution leading to injection of a huge energy density into vacuum. The second reason is related to the inflationary Universe scenario \cite{1}\cite{2}. Inflationary cosmology postulates that there has been a phase of exponential expansion at early Universe driven by the large vacuum energy density of a quantum field (the so-called inflaton). This large value of vacuum density is very important, at least for early versions of inflationary models, and whatever physical process that has led to $\Lambda\approx 0$ today must also allow it to take a large value in the past.\\
Since $\Lambda_{th}$ and $\Lambda_{obs}$ are attributed to different epochs, they are usually measured in different units systems. The quantity $\Lambda_{th}$ is related to early stages of evolution where quantum behaviors of the Universe are important and it is therefore measured in the quantum units system. On the other hand, $\Lambda_{obs}$ is related to the late-time stages where classical properties of the Universe are dominated. In this latter case, physical scales are so large that using the quantum units becomes too cumbersome. In this situation, one usually uses the cosmological units system in which lengths are measured by parsec or light-year rather than Angstrom and time intervals are measured by cosmic year\footnote{Cosmic year (or galactic year) is the time that the sun needs to orbit around the center of the Milky Way galaxy.} or the reciprocal of the Hubble constant ($H_0^{-1}$) rather than period of electron orbits in an atomic system.\\
Let us make a closer look at the issue by a dimensional analysis. By taking $L$, $T$ as basic units for length and time, the dimension of the cosmological constant will be $[\Lambda]=L^{-2}=T^{-2}$. The scale ratio of $\Lambda_{obs}$ and $\Lambda_{th}$ is then given by
\begin{equation}
\frac{\Lambda_{obs}}{\Lambda_{th}}=\left(\frac{T~ (in~the~quantum~ units)}{T~(in~ the~cosmological~ units)}\right)^2
\label{cc}\end{equation}
$$
=\left(\frac{f(\hbar, m_e,...)}{g(G, H_0^{-1},...)}\right)^2\equiv\lambda~~~~~~~~~~~~~
$$
where $f$ and $g$ are some functions. In the usual interpretation, $f$ and $g$ are constants and so is $\lambda$ as the conversion factor relating $\Lambda$ in the two units. The $\Lambda$-problem is the question that why $\lambda$ is so vanishingly small? \\ If some FCs have been varied during expansion of the Universe, then $\lambda$ would be no longer a constant. Note that cosmological scales enlarge with expansion and $\lambda$ would be necessarily a decreasing function of time and acts as a damping factor. Thus variations of FCs during expansion of the Universe automatically provides us with a reduction mechanism for reducing $\Lambda_{th}$.
As the last point, we would like to emphasize that our aim here has been to introduce the basic idea and we have avoided presenting technical matters. There are some questions that we have not concerned in the context of the present work such as: what are the details of the function $\lambda=\lambda(t)$ or what is the field theory which determines the dynamics of $\lambda(t)$? Is the value of $\lambda(t)$ at the present epoch consistent with current cosmological observations? The answers of these and other related questions may be found in \cite{bis}.
~~~~~~~~~~~~~~~~~~~~~~~~~~~~~~~~~~~~~~~~~~~~~~~~~~~~~~~~~~~~~~~~~~~~~~~~~~~~~~~~~~~~~~~~~~~~~~~~~~~~~
\section{Conclusion}
We have briefly reviewed the theoretical reasons for suspecting the constancy of FCs. We then pose the question that how these variations affect conversion factors and units transformations. It is argued that while in human time scales everything remains unchanged, comparing physical quantities, characterizing properties of the Universe, at early and late times may be affected. As a typical example, we have investigated the cosmological constant problem. This problem can be viewed as a controversy between numerical values of $\Lambda$ in  different epochs and units systems. Our analysis establishes a relationship between this controversy and possible variations of FCs. We have considered the possibility that the $\Lambda$-problem be a consequence of regarding naively $\lambda$ as a constant in whole history of the Universe. The present invention provides a reduction mechanism based on dynamics of $\lambda$ and works with expansion of the Universe.

\end{document}